\documentclass[10pt]{article}
\usepackage{sao2}
\usepackage{psfig,epsf}

\issue{2010, 36, 7--13}

\begin{document}
\markboth{O.V.~Verkhodanov, M.L.~Khabibullina}
	  {On the Spectral Index of Distant Radio Galaxies}
\title{On the Spectral Index of Distant Radio Galaxies}
\author{
O.V.~Verkhodanov\inst{a}
\and M.L.~Khabibullina\inst{a}
}
\institute{
\saoname
}
\date{February 26, 2009}{}
\maketitle

\begin{abstract}
The problems of using the spectral index of radio galaxies in various tests, in particular, in
selecting distant radio sources are considered. The history of the question of choosing a criterion of 
searching for distant radio galaxies based on the spectral index is presented. For a new catalog of 2442 radio 
galaxies constructed from NED, SDSS, and CATS data, an analytical form of the spectral index.redshift 
relation has been determined for the first time. The spectral index.angular size and spectral index.flux 
density diagrams have also been constructed. Peculiarities of the distribution of sources on these diagrams 
are discussed. 
\mbox{\hspace{1cm}}

\end{abstract}

\section{INTRODUCTION}

Radio galaxies are among the most powerful observed
 space objects, making it possible to use such 
radio sources to investigate the properties of the Universe
 at various cosmological epochs. Therefore, constructing
 samples of radio galaxies from various red-
shift ranges is one of the most important tasks in 
observational cosmology. 

The sharp fall in the number of observed radio 
sources with increasing redshift forces us to seek 
for ways of rapidly selecting distant objects. One 
of the ways is selection by the spectral index $\alpha<
-1.0$ ($S\sim\nu^{\alpha}$,
where S is the flux density and $\nu$ is
the frequency). It is based on the fact that the farther
 the object, the more probable that it will have 
a steep spectrum. This is one of the first discovered 
and strongest criteria in searching for distant galaxies 
based on radio-astronomical data. It was established 
independently in several works devoted to the identification
 of radio sources and the analysis of radio-
spectra statistics. Among the first papers, we will 
note the paper by Whitfield (1957), who pointed out 
a correlation of the spectral index for a radio source 
with its distance, and the papers by Dagkesamanskii
 (1969, 1970), who found that there were no distant
 objects with spectral indices $\alpha>-0.7$ for the
3C sources. Subsequently, Tielens et al. (1979) determined
 that the fraction of optically identified sources 
with ultrasteep spectra (spectral indices $\alpha^{5000}_{178}<-1$)
decreased with decreasing $\alpha$. Following this work,
Blumenthal and Miley (1979) showed the spectral 
index for 3C and 4C radio sources of various populations
 to depend on the properties of the objects: 
their apparent magnitude, redshift, radio luminosity, 
and angular size. The detected correlation suggested 
that the steep-spectrum sources were, on average, 
further away and more luminous than the sources 
with less steep spectra of the same populations (radio 
galaxies and quasars). Laing and Peacock (1980) 
investigated the relation between radio spectrum and 
radio luminosity for samples of extragalactic sources 
at 178 and 2700 MHz. The spectra were measured for 
the extended regions of the radio sources, which were 
classified by morphological types. At low frequencies, 
the degree of spectral curvature was found to correlate
 with the luminosity for sources with hot spots. 
At high frequencies, the correlation between spectral 
index and luminosity was confirmed. Laing and Peacock
 (1980) also confirmed the existence of a spectral 
index.redshift relation at 178 and 2700 MHz for 
FR II sources (Fanaroff and Riley 1974). At present, 
many groups and authors use such a selection of candidates
 for distant radio galaxies (see, e.g., Chambers 
et al. 1988; Wieringa and Katgert 1991; Soboleva 
and Temirova 1991; R$\ddot{o}$ttgering et al. 1997;
de Breuck
et al. 2000; Pedani 2003; Verkhodanov et al. 2003; 
Gopal-Krishna et al. 2005; Klamer et al. 2006; Bornancini
 et al. 2006; Kopylov et al. 2006). Although 
the criterion based on the spectral index is very efficient,
 its explanation is not yet completely clear 
(De Young 2002). Three main, widely used ideas that 
explain this dependence can be highlighted: 

-- if the spectral steepness for many radio galaxies
increases with frequency, then the spectra of distant 
objects must be steeper than those of near ones because
 of the factor (1+z);

-- the losses due to the Compton scattering of
cosmic microwave background (CMB) photons by 
relativistic electrons grow, since the CMB radiation 
density increases as $(1+z)^4$; these growing losses
lead to an aging of the population of electrons, causing
 a cutoff or an increase in the steepness of the 
high-energy part of the spectrum where the losses are 
greatest; 

-- the selection effect: only bright sources are seen
at high redshifts, and it can then be said that precisely 
these sources exhibit the greatest ``depletion'' in the
high-energy part of the spectrum. 

Klamer et al. (2006) give yet another possible explanation
 for the $\alpha-z$
correlation. They noted that
the steep-spectrum sources are rather rare among the 
nearest FR I radio galaxies, while those with such 
spectra are located in regions with a high baryon 
density. It can then be assumed that if there is evolution
 of the environment of powerful radio galaxies 
reflected in the richness of the surrounding cluster, 
then, on average, the radio galaxies are more likely 
located in regions with a higher ambient density than 
in less dense regions. Hence follows the observed 
redshift dependence of the spectral steepness. This 
can play its role when the gas density as one goes 
into the past increases as $(1+z)^3$, while the injection
of electrons with steeper spectra occurs naturally and 
is described as a function of the redshift in the first-
order Fermi acceleration processes attributable to the 
decreasing expansion velocity of the hot spots in the 
denser and hotter intergalactic medium (Athreya and 
Kapahi 1998). 

Other authors (Kharb et al. 2008), who studied 
a sample of 13 ``powerful classical double'' FR II
radio galaxies, also analyzed this $\alpha-z$
correlation
and pointed out the existence of a redshift dependence 
of the spectral index for both the hot spots and the 
core. From the presence of a correlation between the 
spectral indices for the hot spots and a core with a 
flatter spectrum, Kharb et al. (2007) drew the conclusion
 in favor of choosing the explanation of the effect 
by the electron aging model. 

Although as yet there is no general, common consensus
 on clarifying the nature of the $\alpha-z$
correlation,
 the very fact that this empirical dependence 
exists is important for studies. Almost all of the distant
 radio galaxies found has passed through the 
stage of such a selection. An example of an object 
selected in this way is the radio galaxy that is the red-
shift record-holder with z=5.19 and spectral index
$\alpha=-1.63$ (van Breugel et al. 1999). Another example
 is the object RC 0311+0507 investigated in the 
``Big Trio'' Program (Kopylov et al. 2006). Its redshift

is z=4.514, it has the record radio luminosity for 
radio galaxies with $z>4$, and its spectral index is
$\alpha=-1.33$.

We prepared a sample of distant ($z>0.3$)radio
 galaxies (Khabibullina and Verkhodanov 2009a, 
2009b, 2009c) using the NED\footnote{\tt http://nedwww.ipac.caltech.edu}
 CATS\footnote{\tt http://cats.sao.ru} (Verkhodanov
 et al. 2005), and SDSS\footnote{\tt http://www.sdss.org} (Schneider et al. 2007)
databases for the subsequent application in various 
statistical and cosmological tests (Verkhodanov and 
Parijskij 2003, 2009; Miley and De Breuck 2008), 
in which a large number of objects of the same 
nature is required to carry out a study. To compile 
the primary list, we used the NED database, from 
which we selected objects with the following parameters:
 the redshift ($z>0.3$) and morphological
properties.radio galaxies. The initial list contained 
3364 object. This sample of objects is contaminated 
by objects with incomplete information or by objects
 with different properties. Therefore, the next 
stage was to clean the initial sample of superfluous 
sources. For this purpose, we selected the objects 
that were removed from the primary list (Khabibullina 
and Verkhodanov 2009a): (1) with photometrically 
determined redshifts and (2) with quasar properties 
based on available published data. The final catalog 
contains 2442 sources with spectroscopic redshifts, 
photometric magnitudes, radio flux densities, sizes of 
the radio sources, and radio spectral indices calculated
 from the results of the cross-identification of the 
list of selected radio galaxies with the radio catalogs 
stored in CATS in the frequency range from 30 GHz 
to 325 MHz. By default, the flux densities are given 
on the Baars scale. The data collected in the catalog 
can be used for cosmological luminosity-redshift,
angular size-redshift, and age-redshift tests.

In addition, the statistical diagrams for the parameters
 of radio galaxies and their evolutionary properties
 can be investigated using the cataloged data. 

The goal of this paper is to construct and analyze 
the spectral index-redshift, spectral index-angular
size, and spectral index-flux density diagrams for a
pure sample of radio galaxies, which can be used to 
estimate the redshifts of radio galaxies and to calculate
 their luminosity function. 

\section{CALCULATING THE SPECTRAL INDEX}

To calculate the spectral indices, we performed 
cross-identification in the CATS database with a 
$200'' \times 200''$
identification window. To remove chance 
field radio objects in the specified box, we used a 
technique of data analysis similar to that described 

\begin{figure}[!th]
\centerline{\hbox{
\psfig{figure=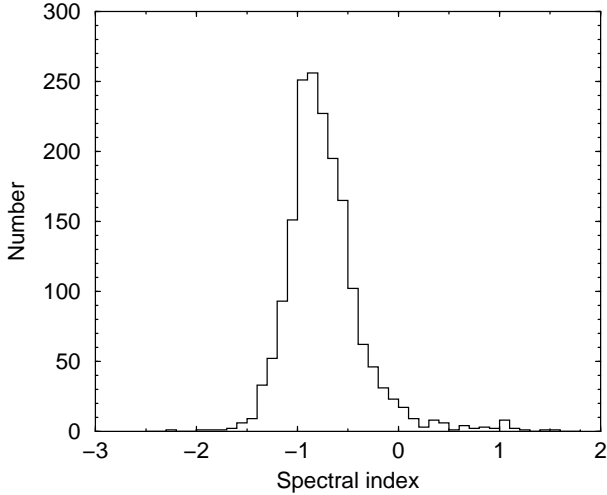,width=8cm,angle=-90}
}}
\caption{Histogram of the spectral-index distribution at
1400 GHz.}
\label{f2}
\end{figure}

in Verkhodanov et al. (2000, 2009). The essence of 
the method is to use a joint analysis of the data in 
coordinate and spectral spaces to separate out the 
probable identifications of specific radio sources at 
various radio frequencies. The spg program (Verkhodanov
 1997a) of the RATAN-600 continuum data 
processing system is used for these purposes. In describing
 the spectra $S(\nu)$ of sources with available
measurements at several frequencies, for the subsequent
 calculation of the spectral indices we used 
an automatic parametrization of $S(\nu)$ by the formula
 $log S(\nu)=A +Bx + Cf(x)$,where S is the
flux density in Jy, x is the logarithm of frequency $\nu$
in MHz, and f(x) is one of the following functions:
exp(-x), exp(x),or $x^2$.For 91% of the sources from
our sample, the spectra are linear. The accuracy of 
determining the coefficients of the least-squares fits 
varies from spectrum to spectrum, but it does not 
exceed 10% of the coefficient. The spectral index at
a given frequency was determined as the slope of the 
tangent to the spectrum at the specified point. 

Figure 1 presents a histogram of the spectral-
index distribution at 1400 MHz. Since the spectra 
of the sample were fitted mainly by linear functions, 
the distribution is virtually independent of the wavelength.
 The median of the spectral-index distribution 
for our entire sample is -0.63. Figure 2 presents the
redshift distribution of objects. 

\begin{figure}[!th]
\centerline{\hbox{
\psfig{figure=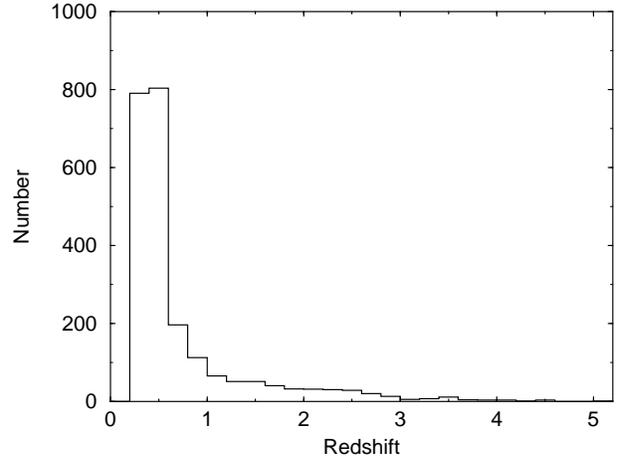,width=8cm,angle=-90}
}}
\caption{
Histogram of the redshift distribution for the catalog.}
\label{f2}
\end{figure}

\section{STATISTICAL DIAGRAMS}

Using the list of selected radio galaxies (Khabibullina
 and Verkhodanov 2009a), we constructed 
the spectral index-redshift $(\alpha-z)$, spectral
index-angular size $(\alpha-\theta)$,
 and spectral index-flux
density $(\alpha-S)$ diagrams.

The spectral index-redshift diagram is shown in
Fig. 3. It has a clear tendency for the spectral index 
to decrease with increasing z and satisfies a linear
 regression $\alpha=a + bz$,where $a=-0.7\pm0.02$
is the regression constant and $b=-0.15\pm0.01$
the slope. Some of the objects have peaks in their 
spectra at a frequency near 1 GHz and lower, giving a 

\begin{figure}[!th]
\centerline{\hbox{
\psfig{figure=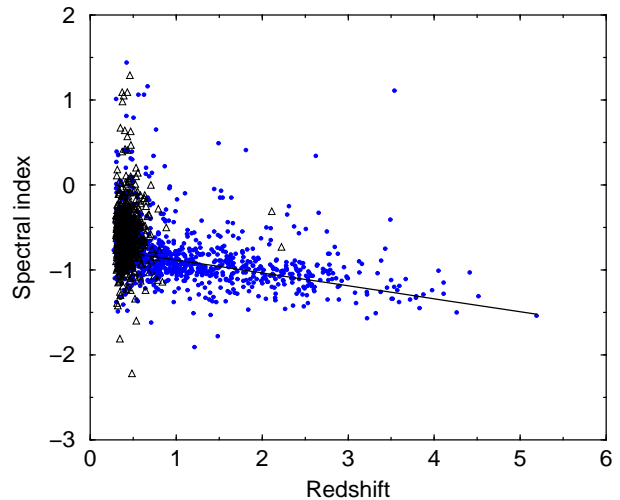,width=8cm,angle=-90}
}}
\caption{
Spectral index-redshift diagram. The spectral
indices were calculated at 1400 MHz. The triangles and 
circles indicate the SDSS objects and the objects from all 
the remaining catalogs, respectively. The regression line 
was drawn using the median spectral indices within bins 
with a step $\Delta z=0.5$.}
\label{f3}
\end{figure}

\begin{figure}[!th]
\centerline{
\psfig{figure=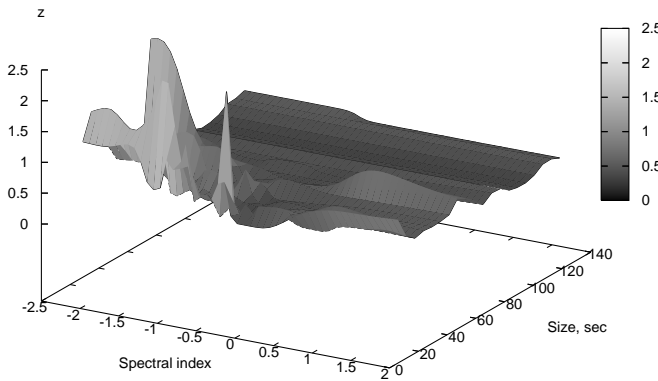,width=10cm,angle=-90}
}
\caption{
Pseudo-three-dimensional angular size-redshift-spectral
index diagram for radio galaxies.}
\label{f4}
\end{figure}

positive spectral index at low frequencies. In the case
of negative $\alpha$, when the source has a steep spectrum,
a straight line was used to fitit. The diagram was 
constructed for the spectral indices at 1400 MHz. 
The presented boundaries of the parameter estimates 
correspond to the statistical ones determined by the 
least-squares method from the median values. The 
accuracy of determining the regression parameters 
can be discussed in terms of sample completeness. 
However, as the above estimates show, the extrapolation
 prediction gives a difference in the right boundary 
value of less than 13% even based on the values
from the first half of the bins. The selection of distant 
objects by the steepness of their spectra has been 
widely used in a number of papers (Blumenthal and 
Miley 1979; Parijskij et al. 1996, 1999; de Breuck 
et al. 2000), but the analytical form of the regression
 for such a large sample of radio galaxies has 
been determined for the first time. The regression 
was calculated from the median spectral indices in 
zintervals with a step (bin) $\Delta z=0.5$. The choice of
the bin size $\Delta z=0.5$ was dictated by two factors: it
should not be (1) large, $\Delta z \leq 1.0$,
in order to take into
account the differential peculiarities in the behavior 
of the population and (2) small, $\Delta z \geq 0.1$,
in order
that there be a sufficient number of objects in the 
specified bins. We chose $\Delta z = 0.5$,on the onehand,
as the mean size and, on the other hand, as a bin that 
allowed a comparatively large number of objects to 
be obtained at $z > 3$. The chosen bin size makes the
regression parameters very stable against an increase 
in the number of objects during a further possible 
expansion of the sample. 

In principle, the existence of a slope can be explained
 by two selection effects, although it does not 
rule out physical causes either. 

(1) The lists of distant radio galaxies were taken 
from the catalogs that were compiled by taking into 
account the selection by a steep spectrum. This effect 
can be avoided only based on complete samples containing
 radio galaxies with measured redshifts. There 
are very few such samples. 
(2) When we choose the most powerful radio 
sources at high z, hot spots with steep spectra 
contribute to the radio emission. The farther the 
radio source, the more probable that it will be more 
powerful than the surrounding ones and will have a 
steep radio spectrum. We are planning to verify this 
fact in our subsequent paper by studying subsamples 
of galaxies with different luminosities. The analytical 
form of the regression can also be used to preselect 
objects at given z, to estimate the distances to radio 
galaxies in the first approximation, and to analyze 
and model the distributions of the number of sources 
(Condon 1984; Gorshkov, 1991), which are used to 
study the luminosity function. The estimate of the 
regression parameters is stable, because it was made 
using the median values of large subsamples of radio 
galaxies. It should be noted that when the spectral 
indices of radio sources from samples including both 
radio galaxies and quasars are analyzed, the regions 
of .with low zare filled (Vollmer et al. 2005). This 
shifts greatly the dependence $\alpha(z)$ toward larger
spectral indices. In addition, to check the stability 
of the slope, we limited the data used by redshift 
intervals with $z < 2.5$ (where more than 15 objects fell
within the bins being analyzed) and reconstructed the 
observed regression for them. As a result, the value 
of .predicted by the regression with incomplete data 
for z=5.2 was -1.34, in contrast to -1.54 obtained
for the complete set of bins, i.e., the relative error in 
the extrapolation to high zis within 13%,whichwe 
consider satisfactory. 

\begin{figure}[!th]
\centerline{
\vbox{
\psfig{figure=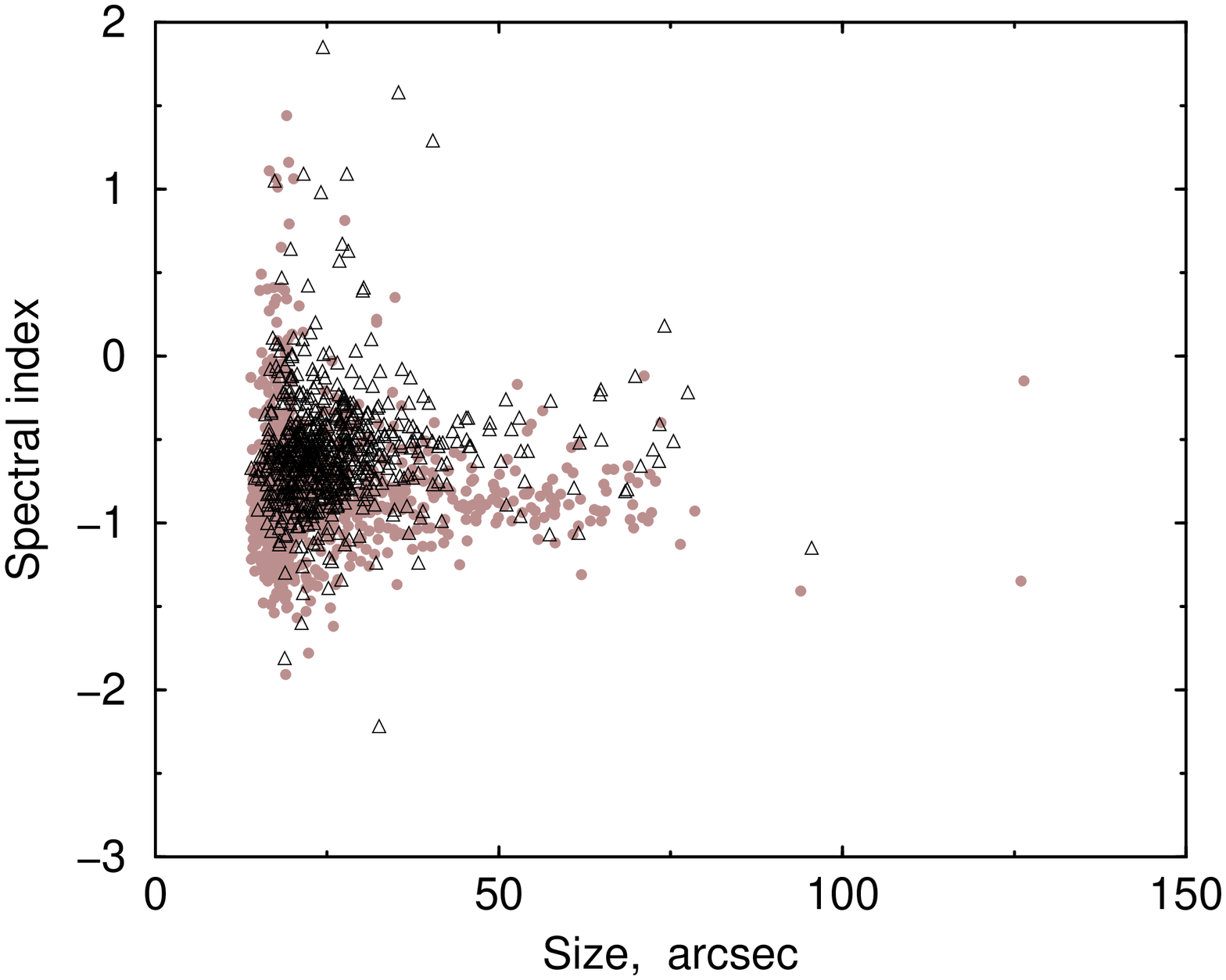,width=8cm,angle=-90}
\psfig{figure=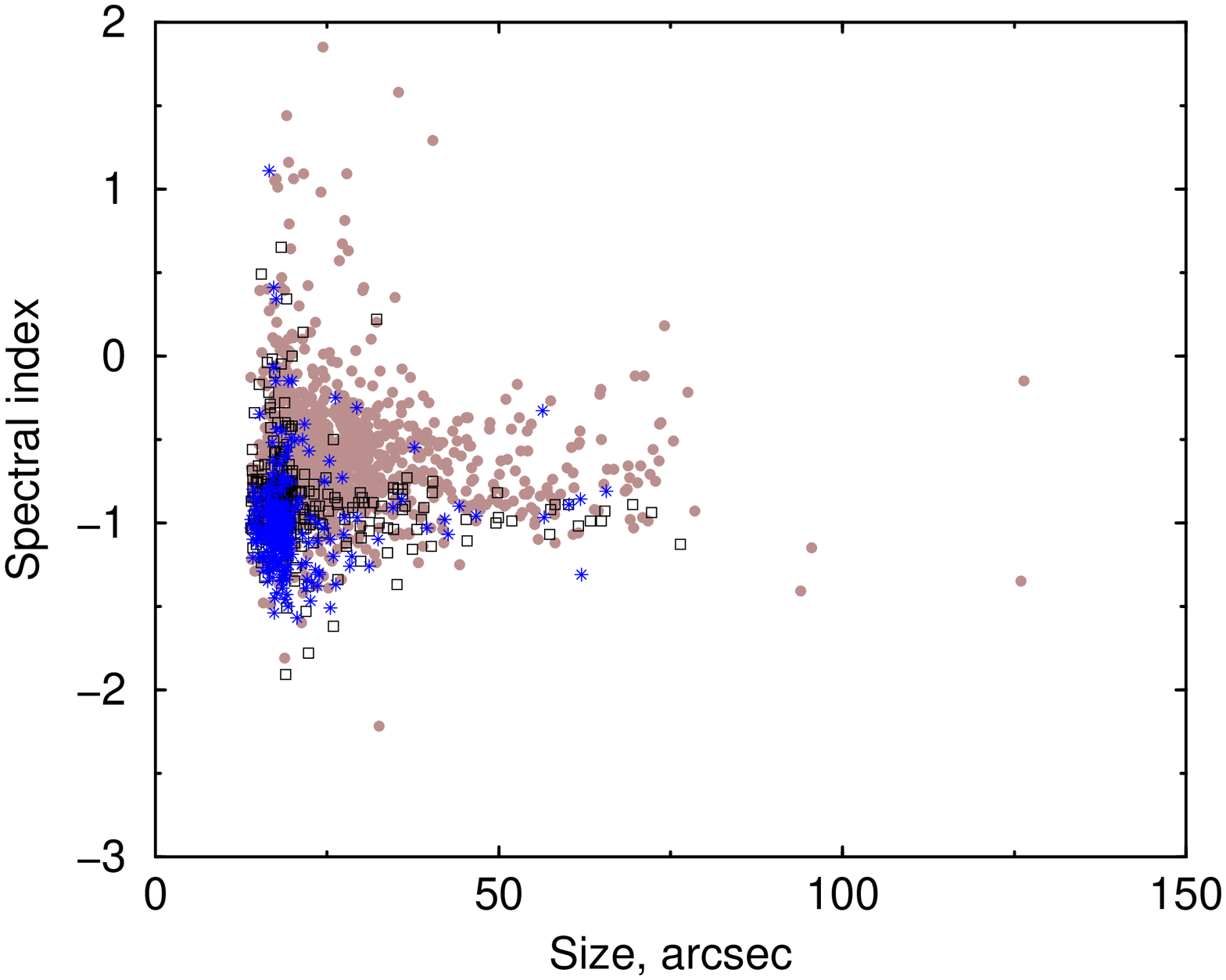,width=8cm,angle=-90}
}}
\caption{
(a) Spectral index-angular size diagram. The
triangles and circles indicate the SDSS objects and 
the objects from all the remaining catalogs, respectively. 
(b) Spectral index-angular size diagram. Differen symbols
 indicate different redshift ranges: $0.3<z<0.7$ (circles),
 $0.7\leq z<1.5$ (squares), and $1.5\leq z$ (asterisks).
The spectral index was calculated at 1400 MHz. The 
measured sizes of the radio sources were taken from the 
NVSS catalog.}
\label{f5}
\end{figure}

\begin{figure}[!th]
\centerline{
\psfig{figure=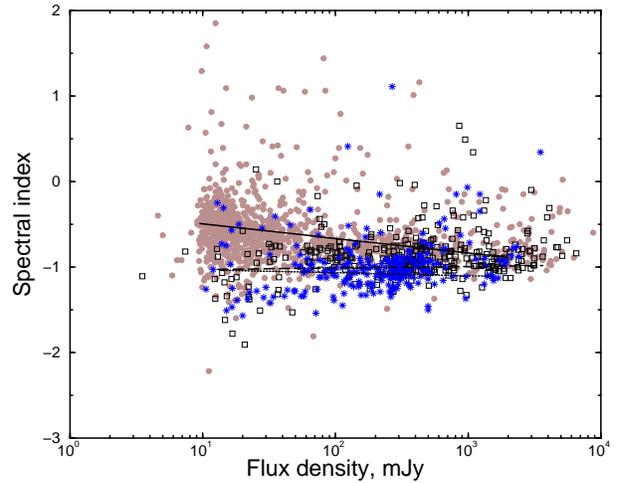,width=8cm,angle=-90}
}
\caption{
Spectral index-flux density diagram for radio
galaxies. The flux density was taken from the NVSS catalog
 at 1400 MHz. Different symbols mark objects from 
different redshift ranges: $0.3<z<0.7$ (circles), $0.7\leq z<1.5$
(squares), and $1.5\leq z$ (asterisks).}
\label{f6}
\end{figure}

For the angular size.redshift.spectral index diagram
 (Fig. 4), we made the spectral index.angular 
size (Fig. 5) and spectral index.flux density (Fig. 6) 
projections; we presented the former in two forms, 
highlighting in one case the SDSS objects by symbols
 and in the other case the objects from different
 redshift ranges by different symbols: $0.3<z<0.7$ (circles),
 $0.7\leq z<1.5$ (squares), and $1.5\leq z$ (asterisks).
The angular sizes were taken from the
NVSS catalog (Condon et al. 1998). The last diagram 
demonstrates the existing relation between the red-
shift of a radio galaxy and the radio spectral index. In 
addition, the expected relation where larger sources 
have steeper spectra due to synchrotron radiation in 
extended components is observed in Fig. 5. Note 
that for the distribution of radio galaxies observed 
in SDSS, there is a shift toward smaller spectral 
indices for the case where the distributions of sources 
in size for the SDSS and non-SDSS subsamples do 
not differ in form. The difference in the data shown 
in Fig. 5 can be explained by the selection effect attributable
 to the existence of a spectral index.redshift 
relation. The latter is demonstrated in Fig. 5b, where 
the region of high zis marked by light gray. 

On the spectral index-flux density diagram
(Fig. 6), we marked galaxies with redshifts in different 
ranges by different symbols: $0.3<z<0.7$ (circles),
$0.7\leq z<1.5$(squares), and $1.5\leq z$ asterisks).
The
flux densities were taken from the NVSS catalog 
(1400 MHz). As we see from the figure, objects 
from all redshift ranges are encountered in the entire 
range of flux densities; most of the objects with 
large spectral indices ($\alpha>-0.3$) are the objects with
the lowest redshifts in our sample. In addition, a 
correlation $\alpha(S_{1400})$ between the spectral index and
flux density is observed for all three zranges. We drew 
linear regressions in the form $\alpha=d+klog S_{1400}$
using the median values of $\alpha$ calculated in logarithmic
 intervals of flux densities. The regressions for 
different ranges are indicated in Fig. 6 by the solid 
line ($d=-0.60$; $k=-1.8 \times 10^{-4}$)
for $0.3<z<0.7$,
the dashed line ($d=-1.03$; $k=-3.7 \times 10^{-5}$)
for $0.7\leq z<1.5$,
and the dotted line ($d=-1.04$; $k=-9.2\times 10^{-5}$)
for $1.5\leq z$.
They confirm the apparent
effect of a change in the slopes of the correlations for 
different redshifts.

\section{CONCLUSIONS}

We presented the results of our investigation of 
the $\alpha -z$, $\alpha - \theta$, and $\alpha -S$
diagrams for a sample
of radio galaxies containing 2442 objects (Khabibullina
 and Verkhodanov 2009a, 2009b, 2009c) and 
constructed from the spectroscopic data collected 
in NED, SDSS, and archives of the Big Trio program 
(Parijskij et al. 1996, 1999). For the objects of our 
list, we found the existence of a regression $\alpha(z)$
and established its analytical form: $\alpha(z)=-0.73-
0.15z$. The estimate of the regression parameters is
stable, since is was made using the median values of 
large subsamples of radio galaxies. This dependence 
can also be used to preselect objects at given z,to 
estimate the distances to radio galaxies in the first 
approximation, and to study the luminosity function. 
Nevertheless, when using the derived relation $\alpha(z)$,
we should take into account the fact that the detected 
regression was obtained from an incomplete sample. 

For the correlation $\alpha(S_{1400})$, the regression parameters
 were found to evolve with increasing z. 
Since correlations (Disney et al. 2008) suggesting 
the existence of a multiparameter fundamental plane 
of radio galaxies are found for a number of their 
physical parameters (total mass, baryon fraction, 
luminosity, etc.), it may be concluded that the evolution
 of the dependence $\alpha(S_{1400})$ is also indicative
of the possible evolution of the fundamental plane. 
In turn, the existence of a correlation between the 
variations in parameters of the fundamental plane 
and spectral indices argues for the explanation of the 
dependence $\alpha(z)$ by the electron aging model, which
complements the conclusions by Kharb et al. (2007). 

\noindent
{\small
{\bf Acknowledgments}.

We wish to thank the referees for useful remarks 
that allowed the content of our paper to be improved. 
In our study, we used the NASA/IPAC Extragalactic 
Database which is operated by the Jet Propulsion 
Laboratory, California Institute of Technology, under
 contract with the National Aeronautics and 
Space Administration. We also used the CATS 
database of radio-astronomical catalogs (Verkhodanov
 et al. 1997, 2005) and the radio-astronomical 
data processing system FADPS\footnote{\tt http://sed.sao.ru/$\sim$vo/fadps\_e.html} (Verkhodanov et al.
1993; Verkhodanov 1997b). This work was supported 
by ``Leading Scientific Schools of Russia'' grants
(S.M. Khaikin.s school) and the Russian Foundation 
for Basic Research (project nos. 09-02-00298 and 
09-02-92659-IND). One of us (O.V.V.) also thanks 
the Russian Foundation for Basic Research (project 
no. 07-02-01417) and the Foundation for Support of 
Russian Science (the ``Young Doctors of the Russian
Academy of Sciences'' Program) for partial support.
}

\end{document}